\documentclass[sigconf]{acmart}
\usepackage{listings}
\usepackage{graphicx}
\usepackage[table,xcdraw]{xcolor}
\usepackage{multirow}
\usepackage{diagbox}
\usepackage{enumitem}
\usepackage{comment}
\usepackage{soul}

\setlist[itemize]{align=parleft,left=0pt..7pt,topsep=0pt}
\lstdefinestyle{verbatimlike}{
basicstyle=\ttfamily\small, 
resetmargins=true,
breakindent=0pt, 
breakautoindent=false,
columns=fullflexible,
frame=single
}

\newif\ifrevision

\newcommand{\textrevision}[1]{%
  \ifrevision
    \textcolor{blue}{#1}%
  \else
    #1%
  \fi
}

\definecolor{shadecolor}{cmyk}{0,0,0,0.15}

\usepackage{graphicx}

\newcommand{\code}[1]{\textnormal{\texttt{#1}}}

\copyrightyear{2026}
\acmYear{2026}
\setcopyright{cc}
\setcctype{by}
\acmConference[ICSE '26]{2026 IEEE/ACM 48th International Conference on Software Engineering}{April 12--18, 2026}{Rio de Janeiro, Brazil}
\acmBooktitle{2026 IEEE/ACM 48th International Conference on Software Engineering (ICSE '26), April 12--18, 2026, Rio de Janeiro, Brazil}
\acmPrice{}
\acmDOI{10.1145/3744916.3787796}
\acmISBN{979-8-4007-2025-3/2026/04}

\title{
Securing the AI Supply Chain: What Can We Learn From Developer-Reported Security Issues and Solutions of AI Projects?
}

\author{The Anh Nguyen}
\affiliation{\institution{School of Computer Science and Information Technology,\\Adelaide University}
\city{Adelaide}
\country{Australia}}
\email{theanh.nguyen@adelaide.edu.au}

\author{Triet Huynh Minh Le}
\affiliation{\institution{School of Computer Science and Information Technology,\\Adelaide University}
\city{Adelaide}
\country{Australia}}
\email{triet.h.le@adelaide.edu.au}

\author{M. Ali Babar}
\affiliation{\institution{School of Computer Science and Information Technology,\\Adelaide University \&\\Elevexai Systems}
\city{Adelaide}
\country{Australia}}
\email{ali.babar@adelaide.edu.au}



\begin{abstract}
The rapid growth of Artificial Intelligence (AI) models and applications has led to an increasingly complex security landscape.
Developers of AI projects must contend not only with traditional software supply chain issues but also with novel, AI-specific security threats.
However, little is known about what security issues are commonly encountered and how they are resolved in practice.
This gap hinders the development of effective security measures for each component of the AI supply chain.
We bridge this gap by conducting an empirical investigation of developer-reported issues and solutions, based on discussions from Hugging Face and GitHub.
To identify security-related discussions, we develop a pipeline that combines keyword matching with an optimal fine-tuned \code{distilBERT} classifier, \textrevision{which achieved the best performance in our} extensive comparison of various deep learning and large language models.
This pipeline produces a dataset of 312,868 security discussions, providing insights into the security reporting practices of AI applications and projects.
We conduct a thematic analysis of 753 posts sampled from our dataset and uncover a fine-grained taxonomy of 32 security issues and 24 solutions across four themes: (1) System and Software, (2) External Tools and Ecosystem, (3) Model, and (4) Data.
We reveal that many security issues arise from the complex dependencies and black-box nature of AI components. Notably, challenges related to Models and Data often lack concrete solutions.
Our insights can offer evidence-based guidance for developers and researchers to address real-world security threats across the AI supply chain.

\end{abstract}

\begin{document}

\maketitle

\section{Introduction}
With the emergence of Artificial Intelligence (AI)-enabled software, developers of these projects face unique security challenges due to the growing complexity of the AI supply chain.
They contend with not only the risks inherited from traditional software, e.g., vulnerable dependencies and insecure coding practices \cite{reichert2024software}, but also novel AI-specific threats such as adversarial manipulations \cite{qin2024adversarial}.
Consequently, multiple high-level risk frameworks, e.g., the OWASP Top 10 for Large Language Models (LLMs) Applications \cite{owasp2025top10}, have been developed.
These frameworks provide valuable guidance on the theoretical dimensions of AI supply chain risks.
However, a significant gap remains between these theoretical frameworks and concrete security problems across different AI supply chain components that developers actually face.
Specifically, it is unclear what security issues are commonly encountered, what challenges they pose, and how developers can effectively mitigate or resolve them.

Discussions on open-source platforms such as Hugging Face (HF) and GitHub (GH) can provide insights into real-world security issues in different components of AI-enabled projects.
These platforms are central hubs where developers not only share models, code, and datasets but also engage in critical discussions that drive innovation and problem-solving \cite{dabbish2012social, jones2024we}.
Security-wise, discussions on these platforms include the security of application, source code, and infrastructure on GH \cite{pletea2014security, buhlmann2022developers, horawalavithana2019mentions}, as well as the security of AI models, e.g., model serialization vulnerabilities on HF \cite{pepe2024hugging}. 
An example of such security-related discussions is the Ultralytics supply chain attack report \cite{ultralytic_attack}, where multiple issues were raised from downstream projects \cite{rten_affect, comfyui_affect, comictranslate_affect} when developers found that the model package was compromised to install crypto miners.
This attack demonstrates how a single vulnerability in the AI supply chain can have widespread consequences.
The community discussions raised during the attack have proven to be critical data sources in understanding the practical security landscape for developing AI-enabled projects.
However, to the best of our knowledge, there is no prior empirical effort to identify and analyze these large-scale discussions to distill real-world security issues that affect the AI supply chain and potential solutions to address these issues.

Our study bridges these gaps by empirically investigating security issues and solutions raised by developers on open-source platforms.
We propose a new pipeline combining keyword matching and an optimal fine-tuned \code{distilBERT} model, selected from a wide range of tested deep learning and large language models, to identify security-related discussions on HF and GH.
Using this pipeline, we have identified a large-scale dataset of 312,868 security-related discussion posts, enabling representative sampling for our subsequent qualitative analysis.
We leverage these posts to characterize security reporting practices in AI-enabled projects.
We then perform a thematic analysis of 753 sampled posts to manually construct a fine-grained taxonomy, capturing real-world AI security issues and corresponding solutions.
These analyses provide synthesized insights to inform the integration of security into the development and maintenance of AI projects.
Our key \textbf{contributions} are:
\begin{itemize}
\item To the best of our knowledge, we present the first empirical study on real-world security issues and solutions in the AI supply chain using developer discussions on HF and GH.
\item We empirically evaluate various DL models and LLMs to select the best classifier (MCC: 0.79) to curate a comprehensive dataset of AI security discussions.
We reveal an exponential growth in security issues that far outpaces formal CVE reporting, and are  more commonly found in text generation and foundation models.
\item We develop a taxonomy of 32 security issues and 24 solutions across AI components (System and Software, External Tools and Ecosystem, Model, and Data).
We find that many AI-specific vulnerabilities stem from the complex dependencies and data flows in AI-enabled projects, with dependency incompatibilities and unsafe deserialization among the most frequent issues.
The black-box nature of models and data further introduces unique security problems, such as malicious output generation and sensitive data leakage.
Critically, we observe a significant gap between identified issues and available solutions in the Models and Data themes, where fixes are scarce and are often indirect.
We also uncover the cross-cutting nature of solutions, as issues in one domain are often resolved by fixes in another.
Overall, our findings highlight the need for a holistic view of the AI supply chain, and
provide recommendations for securing AI-enabled projects.
\item We release our source code, models, datasets as a replication package \cite{replication_package}, and our full taxonomy is publicly available at \cite{tax_viz}.
\end{itemize}

\vspace{-8pt}
\section{Related Work and Motivation}
\label{related}

\vspace{-3pt}
\subsection{Security of the AI supply chain}
Security challenges of the AI supply chain have become increasingly prominent due to the proliferation of reusable components.
While the software supply chain focuses primarily on source code dependencies \cite{reichert2024software}, the AI supply chain encompasses not only software components but also unique AI assets such as models, datasets, and specialized development platforms \cite{ding2025rusty}.
This makes AI systems susceptible not only to traditional vulnerabilities but also to novel attacks, e.g., model poisoning and data manipulations~\cite{casey2024large, zhao2024models, jiang2022empirical, tidjon2022threat, qin2024adversarial}.
All of these problems fall under a broad definition of \textit{AI security issues}, i.e., any adversarial or accidental conditions that undermine the confidentiality, integrity, availability, and, in turn, the trustworthiness of AI assets and systems \cite{gohil2025managing,pispa2024comprehensive,bieringer2024position}.
We adopt this definition consistently throughout this study.

Current research on AI supply chain security, while valuable, offers a limited perspective.
Studies have used static analysis to identify severe vulnerabilities in models \cite{kathikar2023assessing}, highlighted the exploitability of unsafe hosted models \cite{casey2024large}, and demonstrated how model reuse can amplify these risks through model artifacts~\cite{qin2024adversarial, jiang2023empirical}.
While essential for finding technical vulnerabilities, these static analysis and artifact-based approaches fail to capture the practical, day-to-day security issues that developers actually encounter, discuss, and resolve. 
Moreover, analyzing and understanding AI security issues currently depends on high-level security frameworks \cite{weidinger2022taxonomy, cui2024risk, owasp2025top10, mitreatlas}, \textrevision{which are derived from theoretical models in the literature}.
This creates a gap between theoretical risks and real-world developer concerns.
This study fills that critical gap by using developer discussions on community platforms as a direct lens into real-world security challenges and solutions of the AI supply chain.

\vspace{-14pt}
\subsection{Developer discussions on AI security}
\vspace{-3pt}
Open-source platforms, e.g., Hugging Face (HF) and GitHub (GH), are foundational to AI development, serving as registries for pre-trained models, code bases, and problem solving \cite{ding2025rusty, kathikar2023assessing, jiang2023empirical}.
Issues reported on these platforms offer insight into day-to-day challenges in deploying and maintaining AI systems. 
For instance, studies of open-source AI repositories show that developers reported runtime errors, dependencies, and implementation bugs~\cite{yang2023users, jiang2023empirical, buhlmann2022developers}.
While many discussions focus on functional faults, they are also the primary forums where critical, real-world security threats first surface.
Developers often raise concerns and investigate issues within project repositories \cite{liu2025empirical} before issues are formally submitted to vulnerability databases, e.g., NVD \cite{nvd2025datafeeds}.
However, disclosure through such databases can take up to 392 days \cite{rodriguez2018analysis}.
The \textit{Ultralytics} supply chain attack~\cite{hiddenlayer_ultralytics} serves as a compelling example for this scenario.
The attack began when a template injection vulnerability was exploited to publish malicious packages that installed hidden crypto miners.
The timeline \cite{snyk_ultralytics} shows a rapid, community-driven detection effort on GH, where developers began reporting \textit{``PyPI discrepancy''} and questioning the integrity of the package \cite{ultralytic_attack}.
Downstream projects such as \textit{ComfyUI} and \textit{comic-translate} also independently raised issues \cite{rten_affect,comfyui_affect,comictranslate_affect} after detecting crypto mining activity in their applications.
Developers utilized these threads to diagnose the incident, assess its impact, and share mitigation strategies, with the main issue thread accumulating over 150 interactions~\cite{ultralytic_attack}.
This incident illustrates that developer discussions are more than bug trackers, they function as an essential warning system for security incidents across the AI supply chain, offering timely and community-sourced insights into emerging threats.

Although security-related discussions are critical data sources for security research, their identification is non-trivial.
While studies have attempted to detect security reports on GH manually or with keywords \cite{yang2023users, buhlmann2022developers, zahedi2018empirical,le2020puminer}, recent empirical findings suggest that these approaches significantly underestimate both the volume and diversity of security issues~\cite{ghosh2025wasn, zahedi2018empirical}.
We found that in our dataset, only about 1\% of the issues contained explicit security-related keywords. 
However, when we applied a DL model, the proportion rose to 14\% (see Section~\ref{method}), aligning with findings from other ecosystems~\cite{ghosh2025wasn}.
To overcome these limitations, we leverage DL models and LLMs, which have shown superior performance in software vulnerability prediction \cite{yokoyama2024identifying,le2024software,ghosh2025wasn}, to extract a comprehensive corpus of AI security discussions. 
This enriched corpus not only captures a broader set of implicitly stated security issues often missed by keyword-based methods, but it also reduces false positives in sampling for subsequent qualitative analysis of security issues and solutions in AI-enabled projects~\cite{yu2023security, morrison2018identifying}.

\section{Research Questions}
\label{sec:rqs}
To identify and investigate common developer-reported issues and solutions that are related to security in the AI supply chain, we investigate the following Research Questions (RQs):

\textbf{RQ1: What is the extent of security reporting in AI-enabled projects?}
To understand the extent to which security is reported in AI-enabled projects, we first need a reliable method to identify relevant security discussions at scale.
Manual review is infeasible, and keyword-based methods produce biased results with limited coverage \cite{zahedi2018empirical}. 
While DL and LLM-based techniques are effective in other ecosystems (e.g., npm \cite{ghosh2025wasn}), they are underexplored in AI contexts. 
Therefore, RQ1 first investigates how well learning-based techniques identify security-related discussions.
Then, we distill the security reporting patterns and the distribution of these reports across different AI applications and model types.

\textbf{RQ2: What are common AI security issues?}
Current understanding of the AI supply chain security is largely shaped by theoretical frameworks. 
While valuable, these do not necessarily reflect the daily challenges faced by developers. 
RQ2 addresses this gap by building an empirically-grounded taxonomy of security issues in different AI components that are commonly encountered when building real-world AI applications.
These findings are expected to reveal the \textrevision{challenges} in AI supply chain security.

\textbf{RQ3: What are the solutions to these issues?}
Securing AI supply chain requires not only identifying issues but also implementing effective solutions.
Community discussions provide a rich and practical source for these solutions, capturing mitigation strategies, workarounds, and best practices. 
By analyzing these solutions, we can create a parallel taxonomy that not only complements our understanding of problems (RQ2) but also provides actionable insights for practitioners looking to secure their AI projects.

\begin{figure*}[t!]
    \centering
    \includegraphics[width=\textwidth,keepaspectratio]{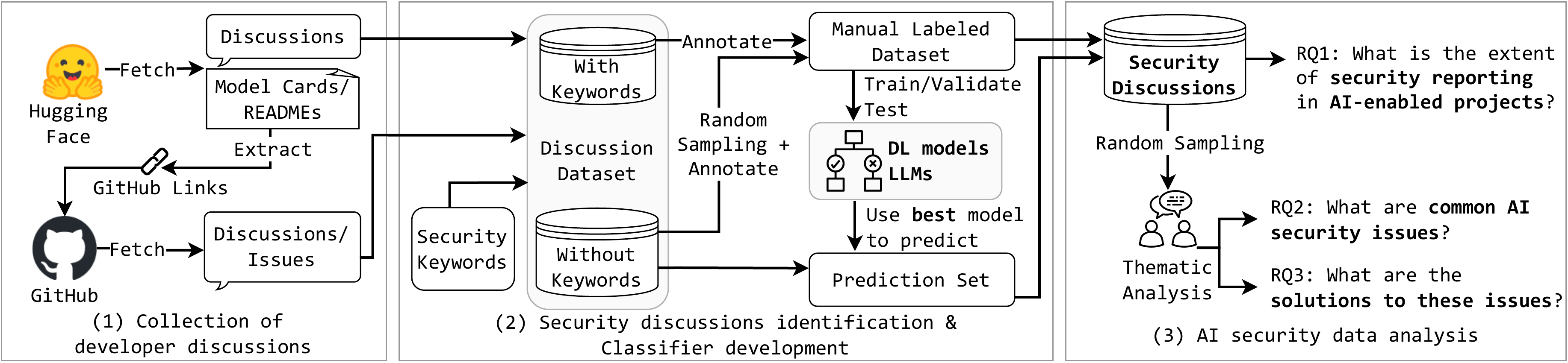}
\vspace{-23pt}
\caption{Research process used to answer the research questions.}
\vspace{-15pt}
\label{fig:method}
\end{figure*}

\vspace{-8pt}
\section{Methodology}
\label{method}
This section describes the methodology we employed. We used a computing cluster with 48 CPU cores, 128 GB of RAM and 2 NVIDIA RTX A6000 GPUs for data collection and model development.

\noindent\textbf{Overview}. As shown in Figure \ref{fig:method}, our research process has three main steps: (\textit{1}) developer discussions  collection, (\textit{2}) security discussions selection, 
and (\textit{3}) data analysis.
First, we crawled all the projects on HF, extracted the GH links, and extracted reported issues and discussions on HF and GH (see Section ~\ref{subsec:data_collection}).
Second, we curated a dataset of security-related discussions, which has two sub-steps.
We first used keyword matching and manual validation to build an initial dataset with high confidence, yet of small size (see Section ~\ref{subsec:kw_matching}).
Then, to increase the coverage of security posts that were missed by keyword and manual searches, we trained and evaluated different deep learning/large language models; the best model was used to detect missing security posts (see Section ~\ref{subsec:classifier}).
Third, we analyzed the extent of security reporting in AI-enabled projects in RQ1, and performed a thematic analysis of a statistically significant subset of all security posts to identify fine-grained AI security issues and solutions in RQs 2 and 3 (see Section \ref{subsec:manual}).

\vspace{-8pt}
\subsection{Collection of developer discussions}
\label{subsec:data_collection}
To focus on AI-enabled projects, we began with HF repositories.
Its curated ecosystem for AI components \cite{ait2025suitability} provides a more relevant starting point than the broader landscape of GH~\cite{kathikar2023assessing}.
Due to the lack of standardized linking between HF and GH \cite{kathikar2023assessing}, 
we identified their connections as follows:
We selected active HF models with at least one ``like'' and one ``download'' to ensure community engagement, which yielded 687,653 models as of January 20, 2025.
We then parsed the model cards and README files for hyperlinks, resulting in 179,045 models with potential links.
Regular expressions were used to extract URLs pointing to GH, yielding 8,972 distinct HF-GH project linkages.
The discussions and issues of these linked projects were then collected from both platforms via their APIs.
Since not all GH repositories enabled the \textit{``discussion''} feature, GH discussions and issues were considered as part of the same GH data source. 
Pull requests and their comments were also included in this source, as the GH API represents both of these as \textit{``issues''}~\cite{github-issues-docs}.
Conversely, we only collected the discussions from HF, as HF repositories did not have a dedicated \textit{``issue''} reporting section.
We then removed 86,421 discussions and issues raised by 36 bot accounts, identified by filtering for account names with the keyword \code{``bot''} and confirming their nature via manual inspection.
This resulted in our dataset with 205,667 HF discussions and 2,022,955 GH issues and discussions.

\vspace{-8pt}
\subsection{Identification of initial security discussions}
\label{subsec:kw_matching}

\begin{table}[]
\centering
\fontsize{8}{9}\selectfont
\resizebox{\columnwidth}{!}{%
\begin{tabular}{lc|c|c|c|c}
\multicolumn{2}{c|}{\textbf{Artifact}} & \textbf{\begin{tabular}[c]{@{}c@{}}All\\ discussions\end{tabular}} & \textbf{\begin{tabular}[c]{@{}c@{}}W/ security\\ keywords\end{tabular}} & \textbf{\begin{tabular}[c]{@{}c@{}}All\\ samples\end{tabular}} & \textbf{\begin{tabular}[c]{@{}c@{}}Security\\ samples\end{tabular}} \\ \hline
\multicolumn{1}{l|}{\multirow{2}{*}{\begin{tabular}[c]{@{}l@{}}\textbf{Train +}\\ \textbf{Val.}\end{tabular}}} & \textbf{HF} & 205,667 & 252 & 252 & 80 \\
\multicolumn{1}{l|}{} & \textbf{GH} & 2,022,955 & 23,962 & 471 & 347 \\ \hline
\multicolumn{1}{l|}{\multirow{2}{*}{\textbf{Test}}} & \textbf{HF} & 35 & 0 & 35 & 7 \\
\multicolumn{1}{l|}{} & \textbf{GH} & 65 & 0 & 65 & 11
\end{tabular}%
}
\caption{Keyword matching and manually labeled dataset from Hugging Face and GitHub. Notes: ``All samples'' refer to the number of records chosen for manual labeling. ``Security samples'' refer to label ``\code{1}'' as described in Sections \ref{subsec:kw_matching}-\ref{subsec:classifier}.}
\label{tab:dataset}
\vspace{-35pt}
\end{table}

To select an initial set of AI security discussions, we first curated a list of popular security keywords, including: ``\code{vulnerability}'', \code{``vulnerabilities''}, ``\code{security}'', \code{``cve''}, \code{``CVE''}, \code{``cwe''} and \code{``CWE''}.
We limited our keywords to only unambiguous security terms and widely used standards (i.e., CVE \cite{cve2024} and CWE \cite{cwe2024}), as we did not want to explicitly target vulnerabilities in a specific area.
The keyword search was then applied to all text fields on the collected dataset, e.g., titles and body text, yielding 252 HF Discussions, 207 GH Discussions, and 23,755 GH Issues.
However, developers usually reported different artifacts (e.g., built-in logs, chat messages) that contained these keywords, but the discussions themselves were not directly related to security.
For example, a developer in \textit{openvinotoolkit/openvino-24013} reported a log file that contained the \textit{``vulnerabilities''} keyword while asking for model serving implementations.
Therefore, we manually labeled all the filtered Discussions from HF and GH, but we manually validated only a subset of 264 GH Issues with at least three keywords, as those with fewer keywords were too numerous for exhaustive manual labeling.
This threshold allowed us to balance coverage with the available manual labeling effort while still capturing a meaningful set of potentially security-relevant issues.
Many of the missing security discussions (e.g., with fewer than three keywords) were expected to be auto-collected using learning-based classifiers in Section~\ref{subsec:classifier}.
We used the label ``\code{1}'' (security-related) for problems where the integrity, confidentiality, or availability of AI models, datasets, and associated software components were affected, potentially leading to security vulnerabilities.
The label ``\code{0}'' (not related to security) was applied to the remaining records.
To minimize subjectivity, the first author independently labeled the whole set and presented the labeling process in frequent meetings with the second and third authors. 
Disagreements were resolved through discussions and mutual agreements, and the process ended when all labels were assigned.

\vspace{-8pt}
\subsection{Security discussions classifier development}
\label{subsec:classifier}
\vspace{-2pt}
To detect missing security posts without security keywords, we selected an optimal classifier from various DL models and LLMs.
The process of preparing the training, validation, and testing data, along with the included models is described hereafter.

\noindent\textbf{Training, validation, and testing data preparation.}
The manually labeled dataset of 252 HF Discussions as well as 471 GH Discussions and Issues in Section \ref{subsec:kw_matching} was shuffled and stratified split by the security label into training and validation subsets with a ratio of 80:20 \textrevision{\cite{nguyen2022vulcurator, yokoyama2024identifying, zhou2021finding, cabrera2021commit2vec, wang2021patchrnn}.} 
This step ensured the ratio of security and non-security records remained roughly the same for both sets.
To ensure that the testing set was not biased toward the keywords in the training data, we further randomly sampled 100 issues that did not contain keywords from all data sources.
After applying the same manual labeling process used while labeling the training/validation sets, we produced a testing set that contained 18 label ``\code{1}'' and 82 label ``\code{0}'' records.
We report both datasets and their labels in Table \ref{tab:dataset}. \textrevision{The datasets, hyperparameters, optimal model settings, and prompts are available in our replication package \cite{replication_package}.}

\noindent\textbf{Deep Learning (DL) models.}
\textrevision{Fine-tuning DL models have demonstrated significant performance for classifying security-related content \cite{ghosh2025wasn,sharma2024distilbert,liu2024spam}.
This process adapts a pre-trained model to our security classification task by training it further on our labeled dataset, allowing it to learn the linguistic patterns and security-related contexts.
The trained model can then predict whether a discussion is security-related.}
We utilized three BERT-based models, including the original \code{BERT} \cite{devlin2019bert}, \code{RoBERTa} \cite{liu2019roberta} and \code{distilBERT} \cite{sanh2019distilbert}, as they have been widely adopted for natural language processing (NLP) and security-related tasks \cite{bharadwaj2022github, minina2023detecting, cao2022sbrpbert, varenov2021security, sharma2024distilbert, wang2021patchrnn}.
We also included two fine-tuned variants, \code{secBERT} \cite{secBERT} and \code{secureBERT} \cite{aghaei2022securebert}, as they have achieved better performance in different tasks with cybersecurity corpus~\cite{aghaei2022securebert} than \code{BERT} and \code{RoBERTa} by leveraging domain-specific data and formal security documents.
We leveraged the original implementations of these models and only modified the last layer to be used as a binary classification head.
The models were then trained with the training/validation set described previously, using the content of the discussions as input text feature, and the security label as the binary classification task label.
To ensure reliable training, we performed stratified 10-fold cross-validation on the training/validation set 
to ensure the security label ratio was roughly the same for each fold, together with early stopping to limit overfitting and randomized grid search to obtain the best hyperparameters for each model.
The final performance was measured with the best-performing hyperparameters using the dedicated testing set described earlier.

\noindent\textbf{Large Language Models (LLMs).}
\textrevision{LLMs can perform classification directly by relying on its pretrained knowledge.
By prompting our task in natural language (zero-shot) and adding examples (few-shot) ~\cite{brown2020language}, LLMs can infer task-specific patterns without additional training, which can be used to predict whether a discussion is security-related.}
We focused our evaluation on notable open-source LLMs to perform the security classification:
\textrevision{\code{llama3} \cite{grattafiori2024llama}, \code{phi4} \cite{abdin2024phi}, \code{mistral} \cite{ollama2023mistral}, and \code{deepseek-r1} \cite{guo2025deepseek}.}
\code{llama3} and \code{phi} are reasoning-focused models, with \code{phi} trained on textbooks and synthetic data.
\code{mistral} and \code{deepseek} aim to balance speed and quality through architectural and training optimizations.
We did not include commercial LLMs due to their black-box nature and potential for unannounced updates, which can lead to changed behaviors~\cite{chen2024chatgpt}.
For the selected LLMs, we employed zero-shot and few-shot prompting~\cite{brown2020language} with a temperature of 0 to obtain deterministic output, and evaluated these models using the testing set described earlier.
In the few-shot setting, we included six examples from our training data, three of which were security-centric to provide sufficient context without overwhelming the prompts and mitigate risks of irrelevant output~\cite{sheng2025llms}.
The prompts were then customized according to the reference from \cite{yokoyama2024identifying}.

\noindent\textbf{Evaluation metrics.} Matthews Correlation Coefficient (MCC) and Macro F1-Score were used to quantify the performance of the classifiers. 
\textrevision{F1-Score, the harmonic mean of precision and recall, balances false positives and false negatives with values ranging from 0 to 1 (best).
MCC, calculated from all four values of the confusion matrix, ensures a more balanced metric and ranges from -1 to 1 (best).
MCC can better reflect performance on imbalanced datasets \cite{nguyen2024automated,le2024mitigating,le2024automatic},} so we reported both but used MCC as the primary indicator.

\begin{figure*}[t!]
\centering  
\includegraphics[width=0.98\textwidth,keepaspectratio]{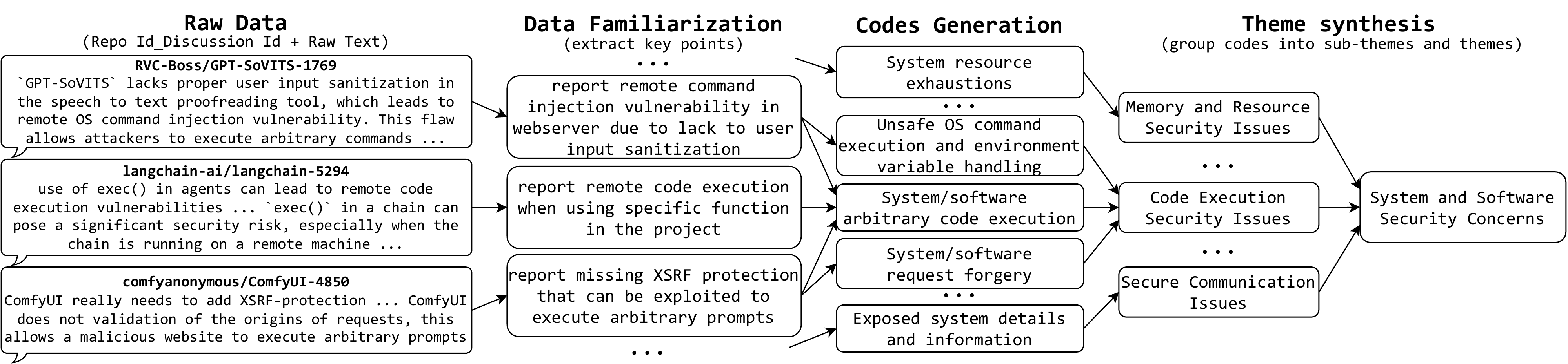}
\vspace{-12pt}
\caption{An illustrative example of the coding process and theme synthesis for RQ2. The same process was applied for RQ3.}
\label{fig:sample_coding}
\vspace{-16pt}
\end{figure*}

\vspace{-8pt}
\subsection{AI security data analysis}
\label{subsec:manual}
\vspace{-2pt}
To perform the analysis for our RQs, we utilized a comprehensive dataset curated from the manually labeled dataset in Section~\ref{subsec:kw_matching} and additional security discussions without security keywords predicted using our best-performing classifier from Section~\ref{subsec:classifier}. 
For \textbf{RQ1}, we examined AI security reporting practices by analyzing their temporal trends, references to official CVE-IDs~\cite{cve2024}, and distribution across AI project domains and model origin types.
These aspects can inform the evolution and the formal reporting process of security issues in the AI supply chain, how they correlate with real-world AI applications and the key practice of model reuse~\mbox{\cite{qin2024adversarial,jiang2023empirical,taraghi2024deep}}.
To analyze the reporting distribution, we obtained the project domains using the \textit{pipeline\_tag} attribute in the model cards \cite{mitchell2019model}, assigning a single task (e.g., \textit{text-generation}) to each project.
These tasks were then grouped under the domains listed on the HF Hub API \cite{huggingface_api}.
For model origin types, we categorized models as either foundational (\textit{base}) or derived (\textit{downstream}) using the \textit{base\_model} attribute \cite{mitchell2019model} and the public count of downstream usages, 
e.g., finetunes~\cite{huggingface_api}.

To address \textbf{RQ2} and \textbf{RQ3}, we employed thematic analysis~\cite{cruzes2011recommended} to analyze the identified security discussions.
This is similar to previous works ~\cite{gao2024documenting, gao2025ai} that studied qualitative attributes of Software Engineering and AI artifacts.
From the inference results and our manually labeled data, we drew a sample size of 368 from HF and 385 from GH. 
\textrevision{These sample sizes ensured statistical representativeness, providing a 95\% confidence level for estimating the proportion of security categories within each source with a 5\% margin of error~\cite{cochran2007sampling}, i.e., the proportions observed in our sample closely approximate those in the entire population.}
\textrevision{During this sampling process, we also kept track of false positive records from our classifier to analyze classifier errors.}
We followed the below three steps to extract data for both research questions simultaneously, where for every discussion, we analyzed both the reported issues and the solutions. 
\textrevision{Following the thematic analysis guidelines~\cite{cruzes2011recommended, braun2006using}, this analysis aims to identify semantic themes, focusing on the explicit content of developers' discussions.
We adopted a realist epistemological orientation, treating the data as a direct reflection of the real-world security challenges that developers face and ensuring our resulting taxonomy is empirically grounded in developer-reported realities.} 

\noindent\textbf{Data familiarization.} The first author collected and examined the documents, then extracted the key points related to security issues and proposed solutions.
During the process, CVE/CWE records were also extracted from NVD \cite{nvd2025datafeeds} if the vulnerability was explicitly mentioned in the issue.
These documents were captured in spreadsheets and shared with the second and third authors for revision.
Disagreements were resolved by discussions and meetings.

\noindent\textrevision{\noindent\textbf{Codes generation and validation.} 
To ensure analytical rigor, we employed an iterative peer debriefing process \cite{lincoln1985naturalistic, creswell2016qualitative} for this step.
Peer debriefing is a collaborative approach where the code book is reviewed by an external party who was not involved in the coding process. 
This method is well-suited for developing and refining a code book through discussion and consensus, which is distinct from other methods that rely on calculating inter-rater reliability against a predefined code book.
Consequently, formal metrics (e.g., Cohen's Kappa) were not calculated, as our goal was collaborative refinement rather than measuring independent agreement \cite{creswell2016qualitative}.
The process started with the first author generating the initial codes by assigning multiple descriptive labels for each sample.}
A set of labels was generated to capture the security issue (RQ2), while a distinctive second set of labels captured the solution or mitigation strategy (RQ3).
A code book was then developed to document the codes, representative samples, and authors' notes for consistent tracking and references.
\textrevision{The second author, uninvolved in the coding, acted as the peer debriefer and reviewed this code book, providing feedback on the clarity, distinctiveness, and representativeness of the codes.}
\textrevision{Regular discussions were held among all three authors to discuss the feedback, resolve disagreements, and iteratively refine the code book by merging, splitting, or renaming codes.}
The process was repeated multiple times until consensus was reached.

\noindent\textbf{Theme synthesis and finalization.}
\textrevision{Following the final code book development, the first author grouped the validated codes into sub-themes and themes that reflect their nature, following thematic analysis standard procedures~\mbox{\cite{braun2006using,cruzes2011recommended}}.}
This synthesis process was also conducted in parallel, with one set of codes organized into the issue taxonomy (RQ2) and the other into the solution taxonomy (RQ3).
Spreadsheets were used to organize the data, key points, codes, and themes to facilitate discussions.
\textrevision{The sub-themes and themes were then reviewed and confirmed by all authors in multiple rounds of discussions, and the process ended when mutual agreement was reached.}
An illustrative example for the coding process from raw data to issue themes (RQ2) is provided in Figure~\ref{fig:sample_coding}, and the same process was also applied for solution themes (RQ3).
The entire process from data familiarization to theme synthesis took approximately a total of 304 man-hours.

\vspace{-6pt}
\section{Results}
\label{results}
\vspace{-2pt}
This section presents the results of the empirical experiments to classify AI security discussions and the respective analyses, as per the research questions in Section \ref{sec:rqs}.

\vspace{-8pt}
\subsection{RQ1: What is the extent of security reporting in AI-enabled projects?}
\label{rq1}
We present and compare the performance of the classifiers described in Section \ref{subsec:classifier} in detecting security-related discussions, and show the security reporting patterns in AI-enabled projects.

\vspace{-6pt}
\subsubsection{\textbf{Performance and comparison of different DL models and LLMs in detecting AI security discussions.}}
As reported in Table \ref{tab:rq1}, \code{distilBERT} was the top performer, achieving the best overall result (F1: 0.89, MCC:0.79) on the combined dataset.
\code{RoBERTa} performed best on the HF subset (F1: \textrevision{0.90}, MCC: \textrevision{0.81}), demonstrating the effectiveness of task-specific fine-tuning for this domain.
Notably, the domain-adapted \code{secBERT} and \code{secureBERT}, while outperforming the baseline \code{BERT}, did not match the performance of \code{distilBERT} or \code{RoBERTa}.
This suggests while security-aware training contributed positively to the security classification performance, it was not sufficient.
The learned representations of \code{secBERT} and \code{secureBERT} may not have aligned with the language and structure of developer security discussions, which include general \textrevision{discourse}, non-standard vulnerability reports, and vague threat descriptions.
In contrast, LLMs performed substantially worse than fine-tuned DL models, with most MCC scores remaining below \textrevision{0.50}.
The best-performing LLM, \code{llama3.3:70b}, achieved an overall MCC of only \textrevision{0.48} on the combined dataset, and \textrevision{0.47} on the HF subset.
On the GH subset, \code{mistral:24b} achieved the best performance with MCC of \textrevision{0.51}.
Moreover, while the performance generally scaled with model size, e.g., \code{deepseek-r1:70b} better than \code{deepseek}-\code{r1:8b}, even larger LLMs could not overcome their lack of domain-specific knowledge and often failed to identify nuanced security discussions.

\noindent\textrevision{\textbf{Error Analysis.}
Our test set ensured the model's actual performance as it was randomly sampled from the entire corpus and reflected real-world scenarios.
To further enhance reliability, we manually validated the classifier's outputs during the random sampling process of our thematic analysis (Section 4.4).
We identified 116 false positives (13.3\%) and revealed two primary error sources:
\textit{(1)} security-related keywords usage in non-security contexts (e.g., \textit{TheBloke}\slash\textit{dolphin-2.5-mixtral-8x7b-GPTQ-1} contains \textit{``Malware api call sequence embedding''} when referring to malware logs extraction),
and \textit{(2)} security-adjacent discussions not describing actual vulnerabilities (e.g., \textit{cpacker/MemGPT-2037} discussed implementing authorization without raising security concerns).
All false positives are available in our replication package \cite{replication_package}.
While our classifier was effective, these limitations highlight the challenge of distinguishing security-specific discussions.
Consequently, RQ1's results are unlikely to be affected given the margin of errors and the false positive rate.
Findings of RQ2 and RQ3 were derived from the analysis after false positives were removed, thus they are not impacted.}

\begin{table}[t]
\fontsize{9}{10}\selectfont
\centering
\resizebox{\columnwidth}{!}{%
\begin{tabular}{l|lll|ll|ll}
\multicolumn{2}{l|}{} & \multicolumn{2}{c|}{GH} & \multicolumn{2}{c|}{HF} & \multicolumn{2}{c}{All} \\
\multicolumn{2}{l|}{\multirow{-2}{*}{\diagbox{\textbf{Model}}{\textbf{Dataset}}}} & F1 & MCC & F1 & MCC & F1 & MCC \\ \hline
 & \multicolumn{1}{l|}{\code{BERT}} & 0.68 & 0.42 & 0.82 & 0.68 & 0.79 & 0.63 \\
 & \multicolumn{1}{l|}{\code{RoBERTa}} & 0.79 & 0.59 & \cellcolor[HTML]{AEAAAA}{\color[HTML]{3166FF} 0.90} & \cellcolor[HTML]{AEAAAA}{\color[HTML]{3166FF} 0.81} & 0.85 & 0.72 \\
 & \multicolumn{1}{l|}{\code{distilBERT}} & \cellcolor[HTML]{AEAAAA}{\color[HTML]{3166FF} 0.90} & \cellcolor[HTML]{AEAAAA}{\color[HTML]{3166FF} 0.82} & 0.87 & 0.77 & \cellcolor[HTML]{AEAAAA}{\color[HTML]{3166FF} 0.89} & \cellcolor[HTML]{AEAAAA}{\color[HTML]{3166FF} 0.79} \\
\multirow{-5}{*}{\rotatebox[origin=c]{90}{DL}} & \multicolumn{1}{l|}{\code{secBERT}} & 0.72 & 0.45 & 0.85 & 0.71 & 0.79 & 0.61 \\
 & \multicolumn{1}{l|}{\code{secureBERT}} & 0.81 & 0.68 & 0.85 & 0.72 & 0.85 & 0.71 \\ \hline
 & \multicolumn{1}{l|}{\code{deepseek-r1:8b}} & 0.65 & 0.30 & 0.67 & 0.34 & 0.59 & 0.20 \\
 & \multicolumn{1}{l|}{\code{llama3.1:8b}} & 0.64 & 0.32 & 0.65 & 0.39 & 0.66 & 0.37 \\
 & \multicolumn{1}{l|}{\code{phi4:14b}} & 0.58 & 0.20 & 0.63 & 0.32 & 0.64 & 0.34 \\
 & \multicolumn{1}{l|}{\code{mistral:24b}} & \cellcolor[HTML]{AEAAAA}{0.73} & \cellcolor[HTML]{AEAAAA}{0.51} & \cellcolor[HTML]{AEAAAA}{0.70} & {0.45} & \cellcolor[HTML]{AEAAAA}{0.71} & {0.46} \\
 & \multicolumn{1}{l|}{\code{deepseek-r1:70b}} & 0.73 & 0.46 & 0.69 & 0.39 &	0.67 & 0.34 \\
\multirow{-6}{*}{\rotatebox[origin=c]{90}{LLM (few-shot)}} & \multicolumn{1}{l|}{\code{llama3.3:70b}} & 0.62 & 0.42 & 0.67 & \cellcolor[HTML]{AEAAAA}{0.47} & 0.68 & \cellcolor[HTML]{AEAAAA}{0.48}
\end{tabular}%
}
\caption{\textrevision{Performance of DL models and LLMs. 
Notes: Grey values are best for each category. 
Blue values are best overall. 
LLMs zero-shot results omitted due to poor performance.}}
\label{tab:rq1}
\vspace{-20pt}
\end{table}

\vspace{-6pt}
\subsubsection{\textbf{Patterns of AI security reporting.}}
\noindent Using our best classifier (\code{distilBERT}), we performed inference on the unlabeled dataset, e.g., discussions without keywords, to detect additional security-related discussions.
This approach identified a significantly larger set of posts than our initial keyword search, yielding 8,421 potential security discussions on HF and 304,002 on GH, which spanned 4,636 unique projects.
Combining these 312,423 records with 445 manually confirmed issues (label ``\code{1}'' in Table~\ref{tab:dataset}), 
we formed a comprehensive corpus of 312,868 records in total.
These findings confirm the importance of using learning-based approaches together with keyword matching to collect AI security discussions.

\begin{figure}[t]  
    \centering
    \includegraphics[trim=2pt 2pt 2pt 2pt,clip,width=\linewidth, keepaspectratio]{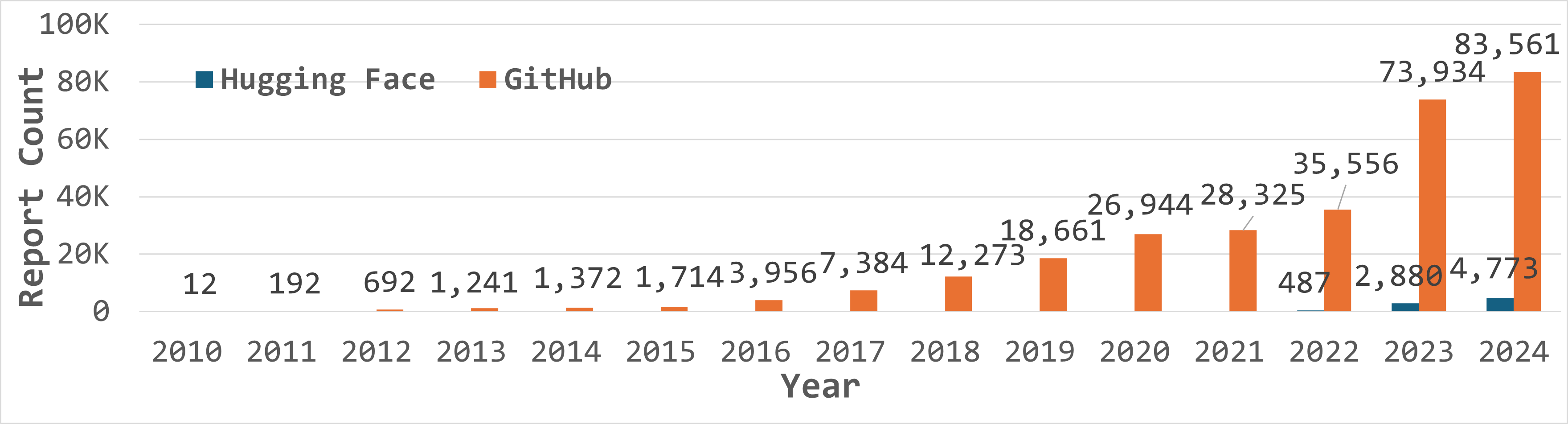}
\vspace{-28pt}
\caption{Number of AI security reports over time.}
\label{fig:rq1_trend}
\vspace{-12pt}
\end{figure}

The temporal distribution of the curated security reports illustrates an exponential increase over the years in AI security discourse (see Figure~\mbox{\ref{fig:rq1_trend}}).
The number of GH security reports more than doubled from 2022 to 2023, and reached a new peak of 83,561 in 2024.
HF Discussion, launched late 2022~\mbox{\cite{hf2022disc}}, shows a similar explosive growth, increasing nearly tenfold from 487 reports in 2022 to 4,773 in 2024.
This rapid growth in community discussions stands in stark contrast to the relatively small number of formally reported vulnerabilities.
Only 320 (< 0.1\%) discussions were linked to CVE-IDs, which are typically required for formal vulnerability reporting.
This highlights the value of platforms like HF and GH in uncovering a broader range of real-world AI security issues that go beyond what is captured through traditional CVE-based tracking.
To further characterize the AI-enabled projects where these security issues commonly arise, we analyzed the distribution of security reporting in terms of project domains and model origin types.

\begin{figure}[t]  
    \centering
    \includegraphics[trim=0cm 9.3cm 19cm 0cm,clip,width=1\linewidth]{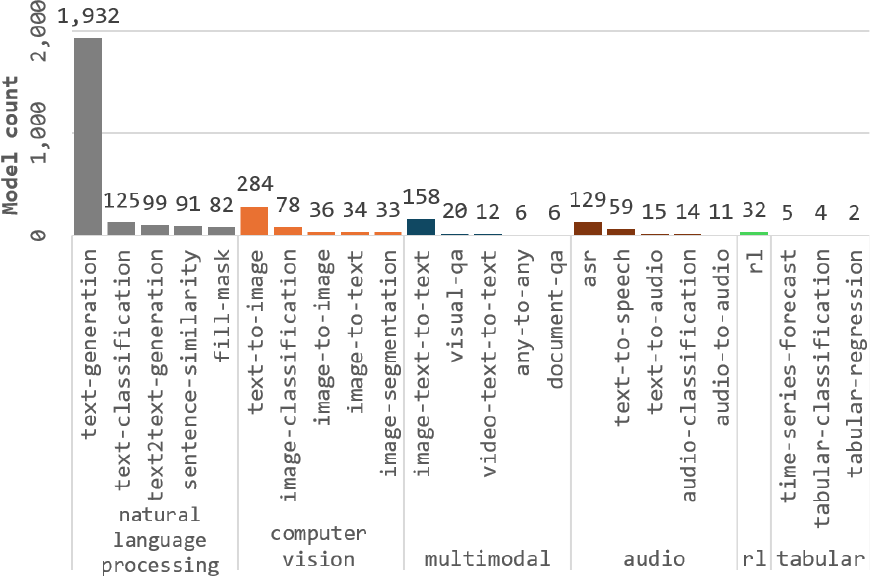}
\vspace{-18pt}
\caption{Distribution of top-5 tasks with potential security-related issues across AI domains. 
Note: asr - automatic speech recognition, rl - reinforcement learning.}
\label{fig:rq1_dist_top5}
\vspace{-12pt}
\end{figure}

\begin{figure}[t]  
    \centering
    \includegraphics[trim=0cm 9cm 0cm 0cm,clip,width=0.8\linewidth,keepaspectratio]{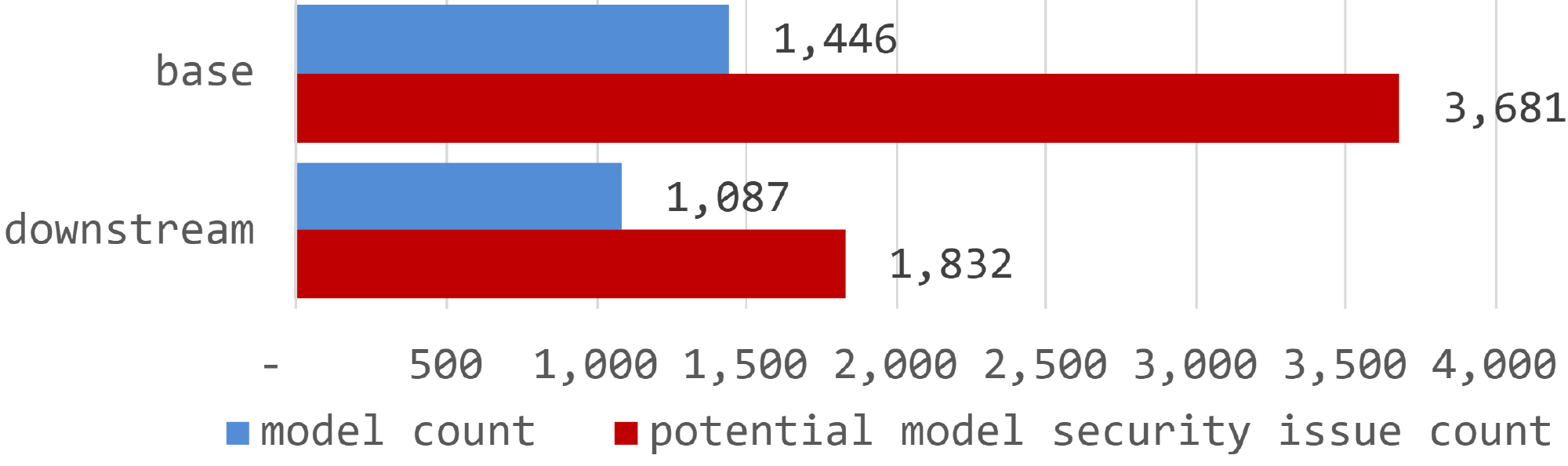}
\vspace{-12pt}
\caption{Distribution of AI model origin types and associated security issues on Hugging Face.}
\label{fig:rq1_dist_lineage}
\vspace{-18pt}
\end{figure}

\noindent\textbf{Distribution of project domains.}
The project domains distribution is reported in Figure \ref{fig:rq1_dist_top5}, in which 950 (20.49\%) models were omitted due to the lack of metadata.
The most prevalent task is \textit{text-generation} in the \textit{NLP} domain with 1,932 models, indicating increased security attention surrounding generative language models.
Other prominent domains include \textit{text-to-image} and \textit{automatic-speech-recognition}, which reflects the broader security implications of generative AI models. 
In contrast, reinforcement learning and tabular data projects showed lower frequencies of possible security discussions, indicating either less risk exposure or under-reporting in developer discussions.
For instance, \textit{keras-io/TF\_Decision\_Trees}, a tabular classification project active since 2022, contained only a single discussion on HF regarding an outdated package, suggesting that security attention is not evenly distributed across AI domains.

\noindent\textbf{Distribution of model origin types.}
Figure \ref{fig:rq1_dist_lineage} reveals that \textit{base} models are associated with a higher average number of security reports (2.55 per model) compared to \textit{downstream} models (1.69 per model).
These numbers suggest that foundation models attracted more security attention, likely due to their wider adoption and foundational role in the ecosystem \cite{taraghi2024deep, qin2024adversarial}.
For example, the base \textit{google/gemma-2b}, widely adopted with 23,536 usages, contained 80 discussions in total, while its fine-tuned \textit{google/gemma-2b-aps-it} variant only contained three discussions on HF.
However, downstream models still exhibit a substantial number of security issues, possibly arising from task-specific adaptations or integrations.

\vspace{-16pt}
\subsection{RQ2: What are common AI security issues?}
\label{rq2}
\vspace{-2pt}
As described in Section~\ref{subsec:manual}, we synthesized 32 issue codes in four main themes: 
(\textit{1}) System and Software, 
(\textit{2}) External Tools and Ecosystem,
(\textit{3}) Model, and 
(\textit{4}) Data.
The first two cover vulnerabilities in code and dependencies, echoing traditional software security but in the AI context; whereas, the \textit{Model} and \textit{Data} themes capture unique AI-specific surfaces, highlighting novel challenges in securing the AI supply chain.
Table \ref{tab:tax} presents our taxonomy with the total ``Frequency'' (the count of all code occurrences for each sub-theme) and the ``Coverage'' (the percentage of unique documents in which the sub-theme appeared at least once).
Below are the codes, sub-themes, and themes in the order of their frequency, along with their description. Note that we only report the key issue codes due to the space limit; the full taxonomy can be found at~\cite{tax_viz}.

\begin{table*}[t]
\centering
\fontsize{12}{13}\selectfont
\resizebox{\textwidth}{!}{%
\begin{tabular}{c||lccc||lccc}
\textbf{Theme} & \multicolumn{1}{c}{\textbf{Issue Sub-theme (RQ2)} } & \# \textbf{Codes} & \textbf{Freq.} & \textbf{Coverage} & \multicolumn{1}{c}{\textbf{Solution Sub-theme (RQ3)}} & \# \textbf{Codes} & \textbf{Freq.} & \textbf{Coverage} \\ \hline

\multirow{5}{*}{\begin{tabular}[c]{@{}c@{}}\textbf{System} \\ \textbf{\&}\\  \textbf{Software}\end{tabular}} 
& System Runtime Compatibility Issues & 3 & 325 & 34.53\% 
& \multirow{5}{*}{\begin{tabular}[l]{@{}l@{}}Secure Coding \& Execution\\Secure Infrastructure \& Deployment \\ Secure Runtime \& Environment Configuration\end{tabular}} 
  & \multirow{5}{*}{\begin{tabular}[c]{@{}c@{}}4\\3\\3\end{tabular}}
  & \multirow{5}{*}{\begin{tabular}[c]{@{}c@{}}153\\124\\92\end{tabular}}
  & \multirow{5}{*}{\begin{tabular}[c]{@{}c@{}}16.47\%\\13.81\%\\11.42\%\end{tabular}} \\
& Code Execution Security Issues & 7 & 229 & 22.58\%   &  &  & \\
& Deployment \& Infrastructure  Security Issues & 3 & 182 & 21.65\% & & & & \\
& Memory \& Resource Security Issues & 2 & 162 & 17.93\% & & & & \\
& Secure Communication Issues & 2 & 47 & 6.24\% &  &  &  &  \\
\hline

\multirow{3}{*}{
\begin{tabular}[c]{@{}c@{}}\textbf{External Tools}\\ \textbf{\&}\\ \textbf{Ecosystem}\end{tabular}
} 
  & \multirow{3}{*}{\begin{tabular}[l]{@{}l@{}}Security of Dependencies Issues \\External Service Usage Security Issues\end{tabular}} 
  & \multirow{3}{*}{\begin{tabular}[c]{@{}c@{}}3\\3\end{tabular}}
  & \multirow{3}{*}{\begin{tabular}[c]{@{}c@{}}201\\138\end{tabular}}
  & \multirow{3}{*}{\begin{tabular}[c]{@{}c@{}}24.04\%\\16.87\%\end{tabular}}
  & Dependency \& Supply Chain Security & 5 & 192 & 23.64\% \\
  & & & & & Security Tooling \& Assurance & 3 & 60 & 7.57\% \\
  & & & & & Platform Identity \& Access Management & 2 & 37 & 4.25\% \\
\hline

\multirow{2}{*}{\textbf{Model}} & Model Output \& Content Control Issues & 3 & 101 & 10.76\% &
\multirow{2}{*}{\begin{tabular}[l]{@{}l@{}}Secure Prompt Techniques/\\Content Safety Filters*\end{tabular}} 
  & \multirow{2}{*}{\begin{tabular}[c]{@{}c@{}}2\end{tabular}}
  & \multirow{2}{*}{\begin{tabular}[c]{@{}c@{}}33\end{tabular}}
  & \multirow{2}{*}{\begin{tabular}[c]{@{}c@{}}4.25\%\end{tabular}} \\
  
& Model Input \& Prompt Security Issues & 2 & 63 & 7.57\% &  &  &  &  \\ \hline

\multirow{2}{*}{\textbf{Data}} & Data Exposure Issues & 2 & 48 & 5.98\% &
\multirow{2}{*}{\begin{tabular}[c]{@{}c@{}}Data Anonymization/Guardrails*\end{tabular}} &
\multirow{2}{*}{\begin{tabular}[c]{@{}c@{}}2\end{tabular}} &
\multirow{2}{*}{\begin{tabular}[c]{@{}c@{}}6\end{tabular}}  &
\multirow{2}{*}{\begin{tabular}[c]{@{}c@{}}0.66\%\end{tabular}} \\
 & Data Integrity Issues & 2 & 29 & 3.72\% &  &  &  &  \\
\end{tabular}%
}
\caption{
Quantitative analysis results of issue (RQ2) and solution (RQ3) sub-themes. 
Notes: The side-by-side layout is for thematic comparison and does not imply a direct mapping. 
(*) The key solution codes in the Model and Data themes are presented instead of synthesized sub-themes due to the low number of samples.
}
\label{tab:tax}
\vspace{-28pt}
\end{table*}

\vspace{-12pt}
\subsubsection{\textbf{System and Software Security Issues (945).}}
\label{rq2_ss}
Since AI-enabled projects contain software components, they are also susceptible to vulnerabilities found in traditional software systems~\cite{papernot2018sok,chen2024security}.
\noindent\textbf{System Runtime Compatibility Issues (325).}
This sub-theme documents how functional conflicts, such as file, framework, or dependency mismatches, can introduce security risks by causing crashes or instability.
Being the most reported sub-theme, incompatibility is illustrated as not only a functional fault but a significant driver of security issues.
For example, \textit{google/long-t5-local-large-1} reported a runtime crash due to an unrecognized file type, while 
\textit{comfyanonymous}\textit{/ComfyUI-5334} documented process failures caused by an unsupported data type in a framework.
In other cases, incompatibilities with the host environment led to severe instability, including \textit{``kernel failure''} (\textit{oobabooga/text-generation-webui-2927}).

\noindent\textbf{Code Execution Security Issues (229).}
This sub-theme captures reports of established code execution vulnerabilities.
Developers have frequently discussed vulnerabilities such as path traversal, XSS, and open redirects, and often linked them to formal CVE reports. 
For instance, \textit{OpenCodeInterpreter/OpenCodeInterpreter-30} identified an \textit{Open Redirect vulnerability} with \textit{CVE-2024-4940}.
A particularly prominent issue, with 85 instances across both platforms, is unsafe file deserialization, which directly leads to arbitrary code execution.
This finding is consistent with existing studies \cite{casey2024large,soodmalicious,zhao2024models} and the official documentation \cite{pythonpickle}.
For example, in \textit{Alpha-VLLM/Lumina-Next-T2I-1}, developers repeatedly warned that using Python's pickle format is \textit{``inherently insecure''} and is a known vector for malware.

\noindent\textbf{Deployment and Infrastructure Security Issues (182).}
This sub-theme includes the failures during deployment.
Insecure default configurations and user misconfigurations~\cite{ye2025llmsecconfig} are most common, where either the default setting is not secure, or users apply settings without understanding the security implications.
For instance, in \textit{huggingface/transformers-22944}, a user criticized the auto-download feature as a \textit{``security hole''} that could download trojan virus, or in \textit{danswer-ai/danswer-2444} where the wildcard CORS setting could lead to information theft.
Another important issue, though less frequently reported, is unaware compromised deployments.
\textit{comfyanonymous/ComfyUI-5165} noted that users blindly using the \textit{``-listen 0.0.0.0''} command had unknowingly exposed \textit{``more than 1,000 instances to the public internet''}, making them vulnerable to complete takeover.

\noindent\textbf{Memory and Resource Security Issues (162).}
This sub-theme covers two distinct types of memory-related risks.
First, resource exhaustion has been frequently reported as a critical availability threat.
Developers in discussions like \textit{microsoft/Phi-3-mini-128k-instruct-46} described persistent memory leaks during model inference, where GPU memory would \textit{``rise until it runs out of memory (OOM).''} 
Second, there have been reports of traditional memory corruption vulnerabilities, such as use-after-free and buffer overflows.
For example, a discussion in \textit{triton-lang/triton-489} pointed to multiple CVEs corresponding to heap-based buffer overflows.

\noindent\textbf{Secure Communication Issues (47).}
This sub-theme highlights two primary failures during communication between system components or services.
The first is information exposure, where systems leak internal details.
For example, a user in \textit{openai/openai-python-1455} warned against exposing server details in query parameters, while another in \textit{ollama/ollama-8020} reported an information exposure issue detected via static scanner.
The second failure is a fundamental lack of secure protocols.
This manifested as either compromised or missing implementations, such as deprecated root certificates \text{threatened the TLS integrity} in \textit{h2oai}/\textit{h2ogpt-1491}, or lacking a \textit{``secure channel''} in \textit{microsoft}/\textit{autogen-4101}.

\vspace{-10pt}
\subsubsection{\textbf{External Tools and Ecosystem Security Issues (339).}}
\noindent This theme captures risks arising from the broader development ecosystem, emphasizing that security of AI projects is heavily influenced by its third-party dependencies, platforms, and tooling.

\noindent\textbf{Security of Dependencies Issues (201).}
This sub-theme concerns security issues of third-party packages.
Developers have reported distinct but related issues: the potential risk of using untrusted sources, exemplified by users questioning the safety of unvetted personal repositories (\textit{TimDettmers/bitsandbytes-1110}); the confirmed presence of actively malicious packages, such as a trojan detected by antivirus in a downloaded dependency (\textit{kohya-ss/sd-scripts-313}); and, most frequently, the use of outdated packages with known vulnerabilities.
The last issue can be exemplified by \textit{abetlen/llama-cpp-python-1711}, where a project's reliance on an old library made it susceptible to an injection vulnerability (CVE-2024-34359), demonstrating the critical need for continuous dependency monitoring.

\noindent\textbf{External Service Usage Security Issues (138).}
Issues related to interactions with external platforms, APIs and tools are captured under this sub-theme.
These include flawed external access control, i.e., authorization and authentication, such as usage of excessive \textit{``write-all permissions''} tokens that created a supply chain attack vector in \textit{keras-team/keras-tuner-929}. 
By compromising a dependency in the pipeline, an attacker could exploit these credentials to take control of the repository and release malicious packages.
Service reliability is another key issue, with outages in external APIs like the HF Inference API causing cascading availability failures for downstream applications (\textit{bigscience/bloom-70}).
Finally, developers have highlighted the limitations of security tools themselves, reporting both false positives, such as scanners incorrectly flagging benign model files in \textit{Mozilla/llava-v1.5-7b-llamafile-7}, and false negatives.
An example of the latter occurred in \textit{protectai/deberta-v3-base-prompt-injection-2}, where prompt injection detectors were easily bypassed with misspelled prompts. 
This shows a critical weakness: even when security tools are in place, their own vulnerabilities can be exploited, creating a false sense of security.

\vspace{-15pt}
\subsubsection{\textbf{Model Security Issues (164).}}
This theme involves vulnerabilities at different stages in the core models unique to AI-enabled projects, i.e., how models process inputs and generate outputs.

\noindent\textbf{Model Output and Content Control Issues (101).}
This sub-theme focuses on issues posed by model-generated content.
\textrevision{A primary issue is the generation of insecure or malicious source code.}
Developers have reported models producing unsafe output, such as a code-generation model that created applications with \textit{``old vulnerable codes''} and contained CVE (\textit{AntonOsika/gpt-engineer-950}). 
In \textit{huggingface}/\textit{autotrain-advanced-848}, models could be forced to \textit{``produce malicious code, including scripts for keyloggers, ransomware, backdoors, trojans''} by manipulating the prompts, highlighting a distinctive model issue that directly creates security risks.
Another issue is the uncontrolled generation leading to denial-of-service, stemming from 1) failure of model generation control parameters, such as incorrect handling of stopping tokens, or 2) inconsistency between components (e.g., tokenizers, models, inference engines), resulting in infinite repetition.
For instance, \textit{mistralai}/\textit{Mistral-7B-Instruct-v0.2-139} documented a model generating \textit{``infinite empty token,''}, resulting in service crash, and \textit{nuprl/MultiPL-E-61} described misconfigured stopping tokens preventing a model from generating functionally correct code.
Finally, we observed issues related to unreliable content filtering mechanisms.
These filters were either too easily bypassed or were over-sensitive, blocking benign output and hinder usability, such as in \textit{microsoft}\textit{/semantic-kernel-10345} where an output filter incorrectly blocked safe generated output.

\noindent\textbf{Model Input and Prompt Security Issues (63).}
This sub-theme relates to how models process user-provided data.
The most frequently discussed issue is model input manipulations and evasions.
This issue includes patterns of prompt injection, where crafted input manipulate the model into unintended actions.
Users in \textit{huggingface}/\textit{tokenizers-1458} reported that prompt templates can be \textit{``exploited by injecting control tokens disguised within Jinja templates to manipulate the AI assistant's behavior, such as redefining system prompts.''}
Beyond text, this code also includes adversarial manipulation of non-text inputs,
with a prime example of the \textit{``image-scaling attacks''} detailed in \textit{tensorflow/tensorflow-62334} where an input image could be manipulated to deceive the model.
Another issue is insecure prompt template design, facilitating injection attacks. 
In \textit{nvidia/Llama3-ChatQA-1.5-70B-1}, developers reported templates that use common strings (\code{`User:'}) as delimiters instead of unique tokens, allowing an attacker to craft inputs that the model misinterprets as a system instruction.
Another example is \textit{turboderp/Mistral-Nemo-Instruct-12B-exl2-1}, where a \textit{``broken tokenizer''} and \textit{``inconsistent prompt formatting''} led to \textit{``persona shifts''} and degraded output.
These inconsistencies can cause reliability and security risks by making the model more susceptible to prompt injection or generating misleading or unauthorized responses \cite{liu2024exploring}.

\vspace{-10pt}
\subsubsection{\textbf{Data Security Issues (77).}}
Another unique aspect of AI-enabled systems is their heavy reliance on a large volume of data that carries their own vulnerabilities~\cite{whang2023data}.
Our sub-themes capture how data can be exposed and how its (low) quality can affect the integrity of a model or a system.

\noindent\textbf{Data Exposure Issues (48).}
This sub-theme highlights risks of either data leakages or opaque data collection.
First, developers have reported risk of models leaking sensitive training or contextual data in their outputs.
In \textit{microsoft/BioGPT-80}, developers found \textit{``response that looks like an artifact of the way training data was formatted''}, potentially exposing sensitive information.
Second, there is a risk of data leakage between isolated components, which mostly stems from architectural flaws where component isolation is insufficient.
In \textit{meta-llama/Llama-3.3-70B-Instruct-52}, developers reported getting \textit{``data from maybe others users as api response, code fragments, chat completion fragments.''}
Similarly, \textit{rocca/openai-clip-js-1} documented an instance of shared local storage can be exploited by \textit{``one app to exfiltrate private data''} of another user.
In addition, external services can also collect data without authorization or data collection policy, as observed in \textit{DS4SD/docling-459} where developers expressed concerns over packages that \textit{``send your documents to some webservice''} without disclosure.

\noindent\textbf{Data Integrity Issues (29).}
This sub-theme focuses on the trustworthiness of data.
A key issue was the risk from deficient data quality, which can lead to data poisoning.
We found reports of training data being contaminated with harmful content. 
For example, a user downloading a dataset in \textit{rom1504/img2dataset-281} reported that it contains \textit{``multiple malicious urls and sites with phishing attacks''}, indicating the training data source itself was a vector for malware.
Poor data quality is also linked to degraded model behavior, such as a model inserting \textit{``unsolicited http-links''} that can lead to phishing sites after being fine-tuned on a poorly processed dataset in \textit{reeducator/vicuna-13b-free-9}.
These issues suggest that low-quality data that have been modified can degrade the integrity of the model, possibly introducing unintended and malicious behaviors.
Another issue is data tampering, where data can be modified without permissions in different environments.
\textit{LAION-AI}/\textit{Open-Assistant-3404} documented this risk as \textit{``possible tampering with data when modifying the container''} when using \textit{``untrusted nodes''}.

\vspace{-10pt}
\subsection{RQ3: What are the solutions to these issues?}
\vspace{-2pt}
This section presents solutions to the identified security issues, ranging from concrete implementations to developer suggestions and best practices.
We have synthesized 24 solution codes in a taxonomy parallel to RQ2 to facilitate a high-level comparison.
Our analysis has revealed that solutions can be cross-cutting, addressing multiple issues across themes.
We first describe the synthesized sub-themes as reported in Table \ref{tab:tax}, followed by the analysis of their relationships.
Similar to RQ2, we only report the key solution codes due to the space limit; the full taxonomy is available at~\cite{tax_viz}.

\vspace{-6pt}
\subsubsection{\textbf{System and Software Security Solutions (369)}}
\label{rq3_ss}
Solutions under this theme apply established software security principles to the code, infrastructure, and runtime environment of AI systems.

\noindent\textbf{Secure Coding and Execution (153).}
These solutions focus on hardening the application source code, mainly to address \textit{Code Execution Issues}.
These include both reactive and proactive input validation, such as replacing insecure functions to prevent path traversal in \textit{microsoft/Olive-83} or proposing new utilities to validate URLs and prevent SSRF (\textit{gradio-app/gradio-8301}). 
A critical solution has been the promotion of secure deserialization practices to mitigate risks from vulnerable formats.
For instance, \textit{huggingface/transformers-27776} proposed gating \code{pickle} usage behind an explicit environment variable and \textit{Alpha-VLLM/Lumina-Next-T2I-1} advised the adoption of the \textit{safetensor} \cite{safetensors} format.
More complex solutions to enable safer execution or implementations like sandboxing (\textit{deepset-ai/haystack-8691}), omitting usage of system-level functions that can execute code (e.g., Python's \code{posix}.\code{system} in \textit{mkiani/gpt2-system-1}), or using time-constant methods to mitigate the potential side-channel attack on password (\textit{cpacker/MemGPT-1230}) were also discussed.
Yet, their adoptions were limited due to the high complexity.

\noindent\textbf{Secure Infrastructure and Deployment (124).}
This sub-theme covers solutions to harden the deployment as a primary approach to address \textit{Deployment and Infrastructure Issues}.
Key strategies include adopting secure deployment practices, such as modifying Dockerfiles to \textit{``run container as non-root user''} to minimize potential cascading risks (\textit{ollama/ollama-8383}).
Other solutions involve (i)  implementing secure-by-design configurations; e.g,. the \textit{coqui-ai/STT-2337} application was designed to require explicit user consent before accessing the microphone to prevent unauthorized recording, and (ii) correcting overly permissive/insecure settings, like removing a wildcard CORS configuration in favor of a strict allow list (\textit{danswer-ai/danswer-2444}).

\noindent\textbf{Secure Runtime and Environment Configuration (92).}
These solutions aim to ensure security and stability during operation, addressing \textit{Memory and Resource}, \textit{Secure Communication} and \textit{System Runtime Compatibility} issues.
To combat resource exhaustion, developers have proposed using different memory management techniques, such as manual garbage collection like \texttt{torch}\allowbreak.\texttt{cuda}\allowbreak.\texttt{empty\_cache()}, to prevent out-of-memory (OOM) errors during long inference tasks (\textit{stabilityai/stable-diffusion-3-medium-91}).
Securing external communication has also been a priority, with solutions ranging from migrating hosts from insecure HTTP to HTTPS (\textit{pytorch/vision-8041}) to correctly configuring system-level certificates to resolve SSL errors (\textit{hiyouga/LLaMA-Efficient-Tuning-5609}).
Lastly, runtime incompatibility issues have been addressed by the modification of runtime configurations, such as modifying size inputs for models and frameworks, enabling specific GPU-driver settings, or installing environment-specific packages (e.g., GPU-enabled vs. CPU-only in \textit{openbmb/MiniCPM-Llama3-V-2\_5-int4-2}).

\subsubsection{\textbf{External Tools and Ecosystem Security Solutions (289).}}
This solution theme manages risks from third-party dependencies, platforms, and tooling used as part of the AI supply chain.
\noindent\textbf{Dependency and Supply Chain Security (192).}
This sub-theme is a major focus, with developers proposing multiple approaches to ensure the security of their dependencies.
The most common solutions involve managing vulnerable dependencies by upgrading, removing, or replacing packages with known CVEs, as discussed in \textit{LAION-AI/Open-Assistant-3546}.
To manage these actions more easily, using dependency management tools, e.g., \code{Poetry} in \textit{infiniflow/ragflow-1257}, has been suggested to resolve dependency conflicts and avoid using outdated packages with known vulnerabilities. 
Beyond managing known vulnerabilities, solutions have also focused on controlling dependency sources by adding checksum and signature validation to installation scripts to prevent tampering (\textit{pixie-io/pixie-243}).
Finally, runtime isolation of dependencies has been suggested as a way to handle conflicts in a secure manner, as documented in \textit{comfyanonymous/ComfyUI-6126}.

\begin{figure*}[t!]
\centering
\includegraphics[width=0.85\linewidth,trim=0cm 8.0cm 0cm 0cm,clip]{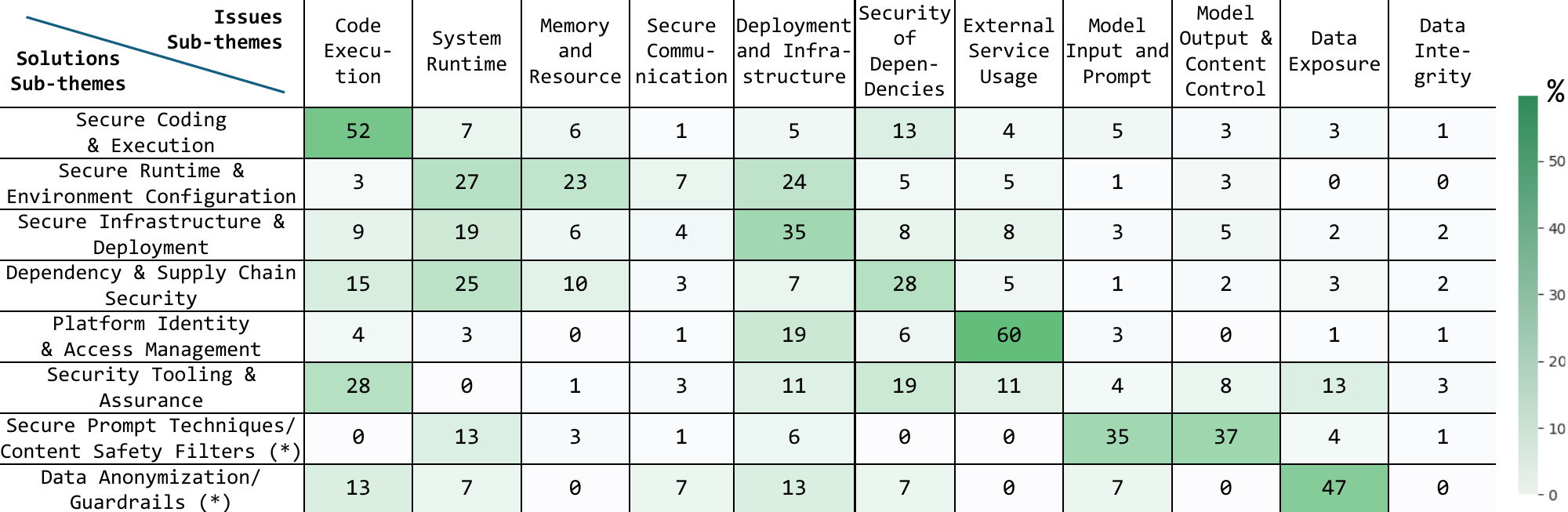}
\caption{\textrevision{Distribution of Issue and Solution Sub-themes.
Note: (*) Solutions in Model and Data are codes instead of sub-themes due to the limited sample size.}}
\label{fig:heatmap}
\vspace{-16pt}
\end{figure*}

\noindent\textbf{Security Tooling and Assurance (60).}
This sub-theme covers the adoption of security-aware processes and tools.
Formal vulnerability disclosure policies (e.g., SECURITY.md) have been adopted, as proposed in \textit{keras-team/keras-tuner-979}, to streamline the security reporting process.
To raise user awareness, developers have added explicit security warnings to the documentation of potentially unsafe functions, such as \code{from\_pretrained()} in \textit{huggingface}/\textit{transformers-18801}. 
Finally, there has been a strong call not only to use but also to improve security scanning tools. 
This includes adding model security  to the project (\textit{TheLastBen/fast-stable-diffusion-525}), and developing evaluation frameworks 
to ensure the reliability of the security tools themselves (\textit{leondz/garak-1104}).

\noindent\textbf{Platform Identity and Access Management (37).}
Solutions in this sub-theme center on implementing robust authentication and authorization, mainly to address the \textit{External Service Usage} issues.
This includes adding authentication measures, such as \textit{``JWT-signed cookie''} in \textit{OpenDevin/OpenDevin-4853}, or enforcing proper authorization and permission controls.
To achieve the latter, developers have proposed detailed plans for fine-grained, role-based access control (RBAC) to restrict sensitive actions like model installation to specific user roles, as noted in \textit{mudler/LocalAI-2102}.

\subsubsection{\textbf{Model Security Solutions (33).}}
\noindent
Solutions in this theme aim to protect model integrity by securing inputs and outputs.
With only 33 samples, we present the codes directly without sub-themes.
On the input side, \textit{Secure Prompt Techniques} focuses on secure prompting, such as optimizing prompt templates or stopping criterias.
\textrevision{
For example, users in \textit{mistralai}/\textit{Mistral-Small-Instruct-2409-10} suggested modifying the Mistral Instruct prompt format to clearly differentiate roles and changing to more robust alternatives such as ChatML. 
In \textit{togethercomputer/RedPajama-INCITE-7B-Instruct-4}, templates with proper newlines 
(\texttt{\textbackslash n\textbackslash n<\textbackslash Details\textbackslash>\textbackslash n\textbackslash n})
were proposed to prevent few-shot examples from leaking into inference results.
In \textit{huggingface/tokenizers-1458}, users suggested inserting \textit{``\texttt{<|system\_prompt|>} at appropriate positions based on the context after tokenization''}
to mitigate the risk of control token injection attacks.
To secure the output, \textit{Content Safety Filters} refers to methods of filtering specific output or restricting the generation of malicious content.
For instance, \textit{langchain-ai/langchain-10441} implemented a tool in the LLM call chain to detect prompt injection and halt model output upon detection.
In \textit{lllyasviel}/\textit{ControlNet-369}, \code{safety}\_\code{checker} was suggested to remove unsafe output, but 
} 
the community has cautioned that this filter alone may be insufficient and require additional safety measures.

\vspace{-8pt}
\subsubsection{\textbf{Data Security Solutions (6).}}
This theme has only six samples, indicating a lack of practical data solutions.
Thus, we present the codes directly.
The identified solutions focus on protecting confidentiality at different stages of the AI lifecycle.
For pre-processing, the \textit{Data Anonymization} solution consists of anonymization techniques to prevent data leakage.
For instance, a combination of automated named entity recognition with rule-based methods was proposed to sanitize sensitive medical texts in \textit{GerMedBERT/medbert-512-3}.
For runtime protection, the \textit{Data Guardrails} solution refers to controlling the output with \textit{``guardrails''} to ensure confidentiality.
In \textit{MaartenGr}/\textit{BERTopic}-\textit{2219}, developers clarified that LLMs do not possess intrinsic security mechanism at training or inference, and developers can \textit{"try to build in guardrails"} with the \code{guardrails} package to protect sensitive information during inference.

\vspace{-8pt}
\subsubsection{\textbf{Cross-cutting nature of solutions and issues.}}
\label{subsubsection:cross}
Our analysis shows that AI supply chain security involves complex, many-to-many relationships between issues and solutions, as illustrated in the issue-solution heatmap (Figure \ref{fig:heatmap}).
While strong and direct correlations exist, such as \textit{External Service Usage} issues being primarily addressed by \textit{Platform Identity and Access Management} through the adoption of authentication and authorization implementations, this is not always the case.
Solutions like \textit{Secure Coding and Execution} and \textit{Security Tooling and Assurance} are versatile and broadly applied to mitigate different issues across themes.
This complexity can be further illustrated by the response of the community in \textit{tensorflow/tensorflow-62334} to the \textit{``image-scaling attacks.''}
The issue itself is related to model, specifically under \textit{Model Input and Prompt Security}, where an adversary deceives the model via its input.
However, the proposed solution was not a model defense but a fundamental software fix, i.e., implementing robust parameter and input validation, which falls under our \textit{Secure Coding and Execution} sub-theme.
The interconnectedness underscores that securing AI systems requires a holistic perspective that goes beyond any individual issue domain.
As a result, implementing an effective AI security posture requires securing multiple surfaces, as a weakness in any one area can be exploited to compromise the entire system.

\vspace{-20pt}
\section{Discussion}
\label{disc}
\subsection{Augmenting existing AI-security knowledge with developer-reported insights}

This section compares our findings with representative AI security frameworks.
The AVID ~\mbox{\cite{avid_ai_tax}} and Cisco's Taxonomy~\mbox{\cite{cisco_ai_tax}} present high-level security issues targeting AI systems, e.g., ``Software Vulnerability'', ``Model Bypass/Compromise'', and ``Data Poisoning'';
whereas, OWASP's Top 10 for LLM Applications~\mbox{\cite{owasp2025top10}} covers simulated/theoretical attack scenarios and respective mitigation strategies.
These frameworks primarily synthesized their knowledge from academic and white papers, which may not reflect actual developer experiences.
Our work enhances these frameworks by highlighting the commonly encountered vulnerabilities and their actual occurrences in real-world AI-enabled projects; many of which were overlooked by existing frameworks.
For example, we uncovered that runtime (in)compatibility between dependencies is the most frequent issue and also unveiled the frequent use of unsafe \mbox{\code{pickle}} format for deserialization.
Such a finer-grained analysis allows for more targeted risk assessment and prioritization based on the component from which a vulnerability originates~\cite{le2022survey,le2022towards}.
Further, we highlighted the current practices and challenges of implementing and adopting solutions, including those proposed in the literature.
For instance, OWASP recommends vetting data and models before use, but we discovered it is not always straightforward to do so in practice due to the usual black-box nature of the data/models.
Overall, our study complements the existing body of knowledge by providing practical insights and lessons learned to address AI security issues.

\subsection{Implications of the findings}
\vspace{-3pt}

\noindent\textbf{Disparity in the security landscape.}
\textrevision{Our results (Table \ref{tab:tax}, Figure \ref{fig:heatmap}) indicate a disparity both in the prevalence and the maturity of solutions between traditional and AI-specific security issues.
AI-enabled projects' heavy reliance on third-party components introduces persistent risks, ranging from system-breaking incompatibilities to inherited vulnerabilities like unsafe deserialization of dangerous file formats.
Yet, security discussions remain dominated by software issues rather than AI-specific concerns regarding Model and Data. 
Given that our dataset includes discussions from both code-focused (GitHub) and model-centric platforms (Hugging Face) and achieved statistical representativeness (Section \ref{subsec:manual}), this prevalence likely reflects broader development practices.
Developers remain primarily concerned with securing foundational software layers, where practical solutions, albeit reactive, have been reported in details, such as configuration management, dependency vetting, and employing security tools.
In contrast, AI's unique themes, e.g., Model and Data, suffer from a profound scarcity of practical solutions both in quantity and quality.}
Although developers have acknowledged security concerns related to model inputs/outputs and data integrity or exposure, available solutions in practice remain limited and rarely discussed or integrated into the existing security routine.
Practitioners often rely on indirect methods or temporary workarounds, such as modifying prompts or inputs without addressing the underlying model vulnerabilities.
These patterns suggest that core AI components are being treated as immutable black boxes, with developers working around security flaws rather than mitigating them from within.
Meanwhile, the research community has proposed proactive model/data-based defense techniques, e.g., adversarial training~\mbox{\cite{zhao2024adversarial}}, membership inference defenses~\mbox{\cite{hu2023defenses}}, and data provenance tracking~\mbox{\cite{padovani2025provenance}}, which directly target these core issues. However, such solutions have yet to be translated into accessible and practical tools for industry use.
This gap highlights a critical need to bridge research and practice by operationalizing these advanced techniques for real-world deployment.
Future work can also explore how traditional and AI-specific problems differ technically, and whether existing security tools can be adapted to address AI-specific threats in projects where both coexist.

\noindent\textbf{Lessons learned for securing the AI supply chain.}
Our results have shown that AI-specific vulnerabilities stem not only from source code and dependencies, but also from non-code artifacts such as data and models.
These AI-specific risks are further heightened by the widespread reuse of critical components, e.g., pre-trained models and public datasets, where developers often inherit weaknesses without full awareness of their provenance, quality, or embedded threats and prior compromises. To ensure the security of the AI supply chain, the full landscape of AI-related risks needs to be clearly documented.
The Artificial Intelligence Bill of Materials (AIBOM)~\cite{nist2024securing} has recently emerged as a promising approach to improving transparency in the AI supply chain. 
However, AIBOM remains in its infancy when it comes to capturing security-related information~\mbox{\cite{xia2024trust,radanliev2024capability}}.
Based on our empirically grounded taxonomy, AIBOM can be extended to capture common AI-specific vulnerabilities and strengthen the security of the AI supply chain.
AIBOM should make model architecture details and known robustness issues of the used model(s) mandatory for any project.
To address data concerns, AIBOM should document data processing steps, data origins, and possible data quality issues along with their mitigation techniques to better inform downstream usage. 
AIBOM should also include information on external platform/component dependencies and disclose any known vulnerabilities associated with them.
Future work can focus on formalizing AIBOM schemas to include these security-relevant metadata, developing automated tools to extract this information from open source platforms, and tracking how the reported security issues evolve over time.
Enriching AIBOM with such practical information would provide transparency into different AI components and help mitigate threats before they are inherited, in turn advancing the security of the AI supply chain.

\vspace{-8pt}
\subsection{Threats to validity}
\vspace{-2pt}

The first threat is the generalization of our study.
\textrevision{Our study focuses exclusively on the open-source AI ecosystem, as developers' discussions are publicly available and provide analyzable records of practical security issues and solutions.
Consequently, our findings may not fully generalize to the challenges faced in closed-source LLMs, which would require different research methods and access to proprietary internal data.}
Within the scope of open-source AI projects, we maximized generalizability by focusing on actively engaged projects and using a learning-based classifier to capture a broad and diverse range of security-related discussions. 
Our subsequent thematic analysis of a statistically significant subset ensures that our taxonomy is representative of common real-world AI security challenges and solutions.

The second threat is the completeness of our analysis due to the reliance on developer-provided metadata.
Due to the low prevalence of explicit links between HF and GH~\cite{kathikar2023assessing}, our dataset might not capture the full population of AI-enabled projects.
Similarly, the analysis of project domains and model origin types was constrained by missing metadata that required us to omit a portion of projects.

The third threat concerns the subjectivity and bias of the manual process employed in our study.
During both the manual labeling and thematic analysis steps, we employed existing processes and validation techniques, e.g., peer debriefing, iterative and frequent discussions, to ensure that the subjectivity was minimized.

\vspace{-5pt}
\section{Conclusion}
\label{conclusion}
We have investigated real-world security issues and solutions in AI-enabled projects with developer discussions on community platforms.
Using a combination of keyword matching and an optimal fine-tuned \code{distilBERT} classifier, we have identified 312,868 security-related discussions on HF and GH.
Our analysis reveals a rising trend in security reports, yet fewer than 0.1\% have official CVE-IDs, highlighting the critical role of these platforms in promptly uncovering real-world AI security issues that may be overlooked by formal CVE tracking.
The identified security issues are particularly concentrated in text generation tasks and are frequently associated with foundation models.
We have also synthesized 32 issue codes and 24 solution codes into four themes: (1) System and Software, (2) External Tools and Ecosystem, (3) Model, and (4) Data.
We have found common security issues across key components of AI-enabled projects and recommended practical solutions to address them. We have also pinpointed areas where solutions are still lacking, which can inspire future work.
Overall, our study takes an important first step toward understanding, identifying, and mitigating real-world security challenges in the AI supply chain.

\balance

\newpage
\bibliographystyle{ACM-Reference-Format}
\bibliography{ref}
\end{document}